\documentstyle{mn}

\newif\ifAMStwofonts

\ifoldfss
  \ifCUPmtlplainloaded \else
    \NewTextAlphabet{textbfit} {cmbxti10} {}
    \NewTextAlphabet{textbfss} {cmssbx10} {}
    \NewMathAlphabet{mathbfit} {cmbxti10} {} 
    \NewMathAlphabet{mathbfss} {cmssbx10} {} 
  \fi
  \ifAMStwofonts
    \ifCUPmtlplainloaded \else
      \NewSymbolFont{upmath} {eurm10}
      \NewSymbolFont{AMSa} {msam10}
      \NewMathSymbol{\upi}     {0}{upmath}{19}
      \NewMathSymbol{\umu}     {0}{upmath}{16}
      \NewMathSymbol{\upartial}{0}{upmath}{40}
      \NewMathSymbol{\leqslant}{3}{AMSa}{36}
      \NewMathSymbol{\geqslant}{3}{AMSa}{3E}

    \fi
  \fi
\fi 

\ifnfssone
  \newmathalphabet{\mathit}
  \addtoversion{normal}{\mathit}{cmr}{m}{it}
  \addtoversion{bold}{\mathit}{cmr}{bx}{it}
  \newmathalphabet{\mathbfit} 
  \addtoversion{normal}{\mathbfit}{cmr}{bx}{it}
  \addtoversion{bold}{\mathbfit}{cmr}{bx}{it}
  \newmathalphabet{\mathbfss} 
  \addtoversion{normal}{\mathbfss}{cmss}{bx}{n}
  \addtoversion{bold}{\mathbfss}{cmss}{bx}{n}
  \ifAMStwofonts
    \ifCUPmtlplainloaded \else
      \UseAMStwoboldmath
      \makeatletter
      \new@mathgroup\upmath@group
      \define@mathgroup\mv@normal\upmath@group{eur}{m}{n}
      \define@mathgroup\mv@bold\upmath@group{eur}{b}{n}
      \edef\UPM{\hexnumber\upmath@group}
      \new@mathgroup\amsa@group
      \define@mathgroup\mv@normal\amsa@group{msa}{m}{n}
      \define@mathgroup\mv@bold\amsa@group{msa}{m}{n}
      \edef\AMSa{\hexnumber\amsa@group}
      \makeatother
      \mathchardef\upi="0\UPM19
      \mathchardef\umu="0\UPM16
      \mathchardef\upartial="0\UPM40
      \mathchardef\leqslant="3\AMSa36
      \mathchardef\geqslant="3\AMSa3E
    \fi
  \fi
\fi 

\ifnfsstwo
  \DeclareMathAlphabet{\mathbfit}{OT1}{cmr}{bx}{it}
  \SetMathAlphabet\mathbfit{bold}{OT1}{cmr}{bx}{it}
  \DeclareMathAlphabet{\mathbfss}{OT1}{cmss}{bx}{n}
  \SetMathAlphabet\mathbfss{bold}{OT1}{cmss}{bx}{n}
  \ifAMStwofonts
    \ifCUPmtlplainloaded \else
      \DeclareSymbolFont{UPM}{U}{eur}{m}{n}
      \SetSymbolFont{UPM}{bold}{U}{eur}{b}{n}
      \DeclareSymbolFont{AMSa}{U}{msa}{m}{n}
      \DeclareMathSymbol{\upi}{0}{UPM}{"19}
      \DeclareMathSymbol{\umu}{0}{UPM}{"16}
      \DeclareMathSymbol{\upartial}{0}{UPM}{"40}
      \DeclareMathSymbol{\leqslant}{3}{AMSa}{"36}
      \DeclareMathSymbol{\geqslant}{3}{AMSa}{"3E}
    \fi
  \fi
\fi 

\ifCUPmtlplainloaded \else
  \ifAMStwofonts \else 
    \def\upi{\pi}
    \def\umu{\mu}
    \def\upartial{\partial}
  \fi
\fi

\title{Gamma rays from interactions of stars with AGN jets}
\author[W. Bednarek and R.J. Protheroe]
       {W. Bednarek$^*$ and R.J. Protheroe \\
Department of Physics and Mathematical Physics,
The University of Adelaide, Adelaide, Australia 5005.\\
$^*$permanent address: Univesity of \L\'od\'z, 90-236\L\'od\'z, 
ul. Pomorska 149/153, Poland.  
               }
\date{University of Adelaide preprint ADP-AT-96-12, submitted to MNRAS}

\pagerange{\pageref{firstpage}--\pageref{lastpage}}
\pubyear{199}

\begin{document}

\maketitle

\label{firstpage}

\begin{abstract}
We have developed a model for gamma ray emission in jets of
active galactic nuclei in which particle acceleration takes place
at a shock in the relativistic jet plasma due to a massive star
in the central region of the host galaxy moving through the jet.
The gamma rays are produced in a pair-Compton cascade in the
radiation field of the star initiated by accelerated
electrons. Our model may account for the observed GeV to TeV
gamma ray spectrum and variability of Markarian 421 and other
blazars detected by the EGRET instrument on the Compton Gamma Ray
Observatory.
\end{abstract}


\section{Introduction}

In recent years many blazar type active galactic nuclei (AGN)
have been detected in high energy $\gamma$-rays by the EGRET
telescope (von Montigny et al. 1995, Thompson et al. 1995).  Of
these, The BL Lac object Markarian 421 has been observed at TeV
energies by the Whipple Observatory (e.g. Punch et al. 1992) and
HEGRA (Petry et al. 1996).  A further BL Lac object not detected
by EGRET, Markarian 501, has also been detected by the Whipple
Observatory (Quinn et al. 1995).  The $\gamma$-ray emission from
blazars is strongly variable on different time scales (from less
than an hour to weeks) in different energy ranges (see Gaidos et
al. 1996, Mattox et al. 1997). In the case of the Markarian 421,
the X-ray variability is correlated with TeV variability
(Takahashi et al. 1996a, 1996b; Buckley et al. 1996) with a time
scale of $\sim1$~week.  Interestingly, the X-ray emission from
this source seems to vary almost periodically on a time scale of
$\sim $1 day (Takahashi et al. 1996a).  Such variability is not
naturally expected in the present popular scenario for the
production of $\gamma$-rays in a blob moving relativisticly along
the jet in a blazar.

Recently, Dar and Laor (1996) proposed that $\gamma$-ray production
in AGN jets may be due to the collision of small clouds with
radii $\sim 10^{12}$ cm and densities $\sim 10^{12}$ cm$^{-3}$
with a highly collimated relativistic proton beam with $\gamma
\sim 10^4$.  In the present paper we consider another scenario as
a possible explanation of collimated $\gamma$-ray production in
blazars.  

In the central regions of active galaxies the processes
of star formation are probably very efficient. For example, in
M32 the central stellar density exceeds $10^7$ M$_\odot$
pc$^{-3}$ (Lauer et al. 1992).  It seems obvious that many stars
must collide with the jet plasma. Here we investigate the
consequences of such frequent stellar wind - jet plasma
collisions for the production of highly collimated $\gamma$-ray
beams in blazars.  The general scenario is illustrated by
Fig.~1(a).

\begin{figure*}
\vspace{6cm} 
\includegraphics{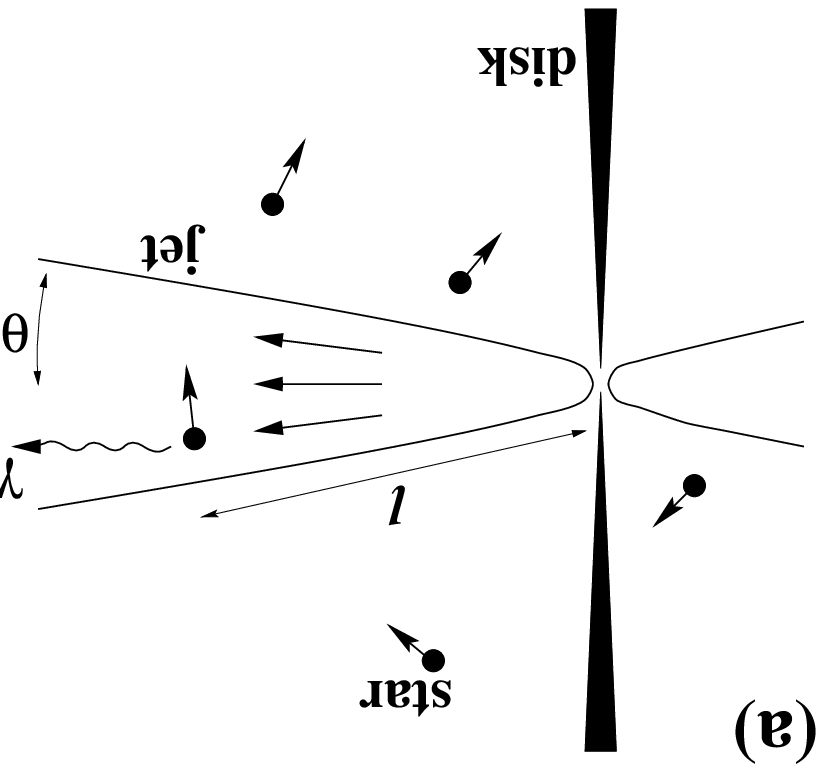}
\includegraphics{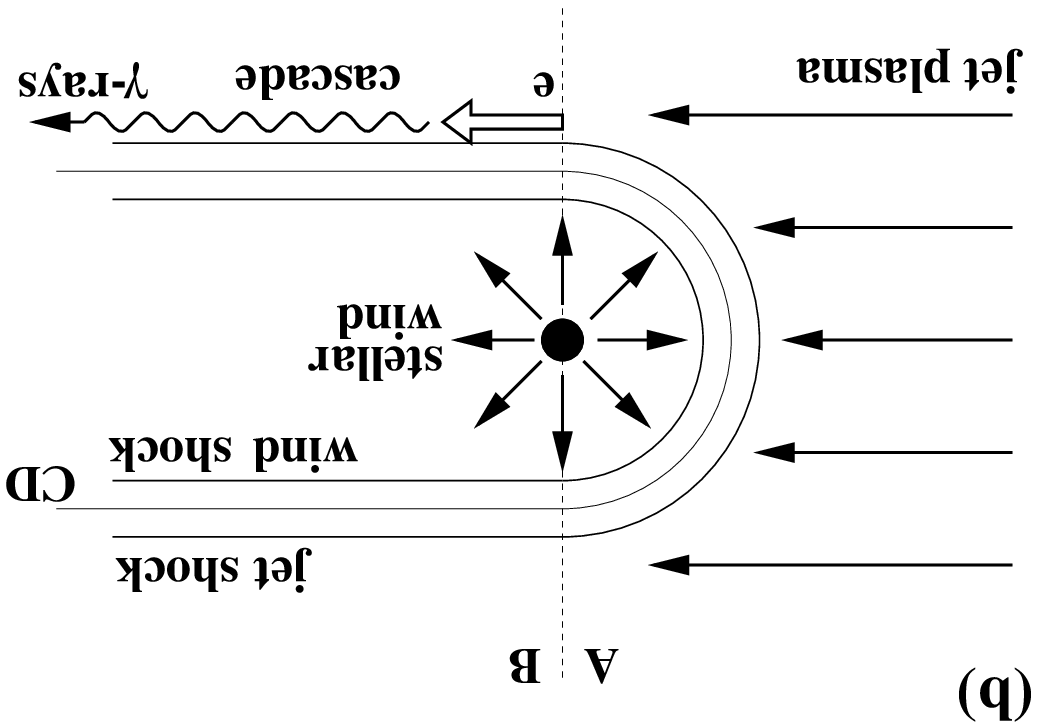} 
\caption{Schematic representation of the
interaction of a star with an AGN jet. (a) The geometry (not to
scale) showing stars orbiting around the centre of the galaxy and
showing a star entering the jet, having opening angle $\theta$, a
distance $l$ above the central engine.  (b) Expanded view of the
region around the star showing the stellar wind and the
relativistic jet plasma interacting to produce a double shock and
contact discontinuity (CD).  In region A the jet plasma shock is
quasi-parallel, and in region B is quasi-perpendicular.
Particles are accelerated efficiently at the relativistic shock
in the jet plasma, and in region B may be highly collimated along
the magnetic field direction which is expected to be parallel to
the direction of jet plasma flow (see text).}
\label{fig1}
\end{figure*}

\section{Star colliding with jet plasma} 

Let us consider a very simple scenario in which a powerful jet,
propagating from the central engine of active galactic nuclei,
meets a massive star. The jet may consist of relativistic
electron-proton (neutral) plasma with power $L_j = L_p + L_e
\approx \dot{N_p} \gamma (m_p + m_e) c^2$, where $\dot{N_p}$ is
the rate of injection of protons with Lorentz factor $\gamma
\sim 10$ into the jet, and $m_p$, $m_e$ are the proton and
electron masses.  The kinetic powers of jets in radio-loud AGNs
may be up to at least $\sim 10^{47}$ erg s$^{-1}$ (Rawlings \&
Saunders 1991, Celloti \& Fabian 1993, Falcke \& Biermann 1995).
Assuming the jet propagates within a cone with opening angle
$\theta$, at a distance $l$ along the jet the ram pressure of the
jet plasma is
\begin{eqnarray}
P_j = L_j/\pi c \theta^2 l^2 \cong 1.5 L_{46} \theta_5^{-2} l_1^{-2} \;\;\; 
{\rm~erg \;cm}^{-3},
\label{eq1}
\end{eqnarray} 
\noindent
where $L_{46}$ is the jet power in units of $10^{46}$ erg
s$^{-1}$, $\theta_5$ is its opening angle in units of $5^\circ$,
and $l_1$ is the distance in parsecs.
  
Stars in the central regions of galaxies move with high
velocities, of the order of $(3- 6)\times 10^3$ km s$^{-1}$, and
are characterised by very intense winds and strong surface
magnetic fields. For example the mass loss rate of Wolf-Rayet
(WR) stars and the terminal velocity of the matter outflow are
$\dot{M}_{\rm WR}\sim (0.8- 8)\times 10^{-5}\dot{M}_{\odot}$
yr$^{-1}$, and $v_{\infty}^{\rm WR}\sim (1- 5)\times 10^3$ km
s$^{-1}$ (Lang 1991). Young massive OB stars also have intense
stellar winds with $\dot{M}_{\rm OB}\sim 10^{-6} \dot{M}_{\odot}$
yr$^{-1}$, and $v_{\infty}^{\rm OB}\sim (1- 3)\times 10^3$ km
s$^{-1}$ (e.g. Lang 1991).

Some of these massive stars will obviously cross the jet cone
from time to time. When this happens, the interaction of
relativistic plasma in the jet with the stellar wind (and the
associated stellar magnetosphere) will result in the formation of
shock waves similar to those discussed in the context of
collisions of winds in binary systems involving early type stars
(e.g. Eichler \& Usov 1993).  In the present context, the
location of the shocks is determined by the parameters of the jet
and the star. A double shock structure, with a contact
discontinuity between them, should form (see Fig.~1b).  The shock
in the stellar wind will be non-relativistic while that in
the jet plasma will be relativistic.

In principle the jet pressure can be balanced by the stellar wind
pressure or the stellar magnetic field pressure. 
For stars with a dipole magnetic field and a strong wind, 
the structure and strength of the magnetic field around the star is
(Weber \& Davis 1967, Usov \& Melrose 1992)
\begin{eqnarray}
B(r) \approx B_s\times \cases { (R/r)^3, & for $ R < r < R\xi$ (dipole), \cr
                              \xi^{-1} (R/r)^2, & for  
R$\xi < r < R \eta^{-1}$ (radial), \cr
                              \eta \xi^{-1} (R/r), & for  $r > R\eta^{-1}$ 
(toroidal).\cr}
\label{eq2}
\end{eqnarray}
\noindent
where $\xi \equiv r_A/R$, $r_A$ is the Alfven radius which
depends on the star's parameters and is in the range
$\sim$(1-3)R, where R is the radius of the star (Usov \& Melrose
1992), and $\eta \equiv v_{\rm rot}/v_{\infty}$ where $v_{\rm
rot}$ is the surface rotational velocity of the star.  Typically
for early-type stars $v_{\rm rot}\cong (0.1- 0.2) v_{\infty}$
(e.g. Conti \& Ebbets 1977).

If the stellar wind is very strong, the jet pressure is balanced
by the wind pressure,
\begin{eqnarray}
P_{\rm wind} = {{\dot{M} v_{\infty}}\over{4\pi r^2}} 
\approx 1.6\times 10^4 \dot{M}_{-5} v_3 R_{12}^{-2}
(r/R)^{-2} {\rm erg~cm}^{-3},
\label{eq3}
\end{eqnarray} 
\noindent
where the mass loss rate of the star is $\dot{M} = 10^{-5}
\dot{M}_{-5} M_{\odot}$ yr$^{-1}$, the wind velocity is
$v_{\infty} = 3\times 10^{8} v_3$ cm s$^{-1}$, and the radius of
a star is $R = 10^{12} R_{12}$ cm.  The double shock structure is
then located at a distance $r_{\rm sh}$ from the star given by
\begin{eqnarray}
r_{\rm sh}/R\approx 103 \dot{M}_{-5}^{1/2} v_3^{1/2} R_{12}^{-1} 
\theta_5 l_1 L_{46}^{-1/2}.
\label{eq4}
\end{eqnarray}
\noindent
The maximum power which may in principle be extracted from the
jet by the shock can be estimated from
\begin{eqnarray}
L_{\rm sh}\cong L_j (r_{\rm sh}/\theta l)^2 
\approx 1.5 \times 10^{39} \dot{M}_{-5} v_3 \;\;\;\; {\rm erg \; s^{-1}}.
\label{eq5}
\end{eqnarray}
\noindent
Note that this depends only on the stellar wind power, but not on
the jet power or distance from the central engine.

If the jet pressure can not be balanced by the stellar wind
pressure, it may be balanced by the pressure associated with the
star's magnetic field, $B(r)^2/8\pi$, provided
\begin{eqnarray}
B_3 > 0.6 \dot{M}_{-5}^{1/2} v_3^{1/2} R_{12}^{-1} 
\label{eq6}
\end{eqnarray}
where $B_3$ is $B_s$ in units of $10^3$ G.
Then the shock location would be given by
\begin{eqnarray}
r_{\rm sh}/R \approx 5.5 (B_3 \theta_5 l_1 L_{46}^{-1/2})^{1/3}.
\label{eq7}
\end{eqnarray}
\noindent
In this case, the maximum power that can be extracted by the shock is
\begin{eqnarray}
L_{\rm sh}\approx 4.4\times 10^{36} R_{12}^2 
(B_3 L_{46} \theta_5 ^{-2} l_1^{-2})^{2/3}.
\label{eq8}
\end{eqnarray}

In the most extreme case of the jet pressure dominating over 
the wind and magnetic pressures of the
star, the jet collides directly with the stellar atmosphere. 
The jet pressure dominates over the magnetic pressure if
\begin{eqnarray}
L_{46} > 2.7\times 10^4 B_3^2 \theta_5^2 l_1^2,
\label{eq9}
\end{eqnarray}
\noindent
and the jet pressure dominates over the wind pressure if
\begin{eqnarray}
L_{46} > 10^4 \dot{M}_{-5} v_3 \theta_5^2 l_1^2 R_{12}^{-2}.
\label{eq10}
\end{eqnarray}
\noindent
If both conditions are met, 
the jet initiates a very strong outflow of matter from the star, which 
pressure balances the jet pressure very close to the star's surface.  

\section{Acceleration of particles}

Electrons and protons are expected to be accelerated by the first
order Fermi acceleration mechanism (e.g. see reviews by Blandford
\& Eichler 1987, Jones \& Ellison 1991). If the diffusion
coefficient is proportional to energy we can write the
acceleration rate as
\begin{equation}
\dot{E} = \chi Ze c B \;\;\; \rm erg \; s^{-1}
\label{eq14}
\end{equation}
where $B$ is in gauss, $Ze$ is in statcoulombs, and $\chi$
depends on the details of the acceleration mechanism.  For
acceleration at a shock with velocity $0.1 c$ values of
$\chi$ as high as 0.04 or $1.6 \times 10^{-4}$ are possible if
the shock is perpendicular or parallel respectively
(e.g. Protheroe 1997).  

For typical parameters of massive stars (surface temperature of
the order of a few $10^4$ K, and other parameters mentioned
above, Lang 1991), accelerated protons are not expected to encounter
sufficient target matter or radiation for interactions in the
shock region.  The interaction lengths (in cm) for protons in the
wind plasma and stellar radiation are approximately
\begin{eqnarray}
\lambda_{pp} \approx 3\times 10^{14} \dot{M}_{-5}^{-1} v_3 R_{12}^2 
(r_{\rm sh}/R)^2,
\label{eq11}
\end{eqnarray}
\noindent
\begin{eqnarray}
\lambda_{p\gamma}\approx 3.4\times 10^{14} T_{4}^{-3} (r_{\rm sh}/R)^2, 
\label{eq12}
\end{eqnarray} 
\noindent
where the star's surface temperature is taken to be $T = 10^4
T_4$ K.  These interaction lengths are longer than the
characteristic distance scale, i.e. the shock radius.  We
therefore concentrate on electron acceleration in this paper.
Using the standard formula for the synchrotron energy loss rate
one obtains, in the absence of other losses, the maximum electron
energy
\begin{equation}
E_e^{\rm max} = 6 \times 10^4 \chi^{1/2} B^{-1/2} \;\; \rm GeV.
\label{eq15}
\end{equation}

As shown in Section~2, collisions of stars with different
parameters ($\dot{M}, v_{\infty}, B$, $l$) with jets having
different parameters ($L_j, \theta$) will result in the formation
of shocks at various distances from the star.  Since the shock
parameters will be different for the shock in the stellar wind
and the shock in the jet plasma, which we shall refer to as ``the
wind shock'' and ``the jet shock'', we shall discuss them
separately below.

\subsection{Shock in the jet plasma}

The jet shock may be very effective for particle acceleration
because it occurs in the relativistic jet plasma.  Observations
of the inner parts of strong jets show that the magnetic field is
mainly directed along the motion of the jet plasma (Saikia \&
Salter 1988).  In such cases the region of the shock nearest to
the central engine (region A to the left of the dashed line in
Fig.~1) is quasi-parallel, while the region of the shock farthest
from the central engine (region B in Fig.~1) is
quasi-perpendicular.  The strength of the longitudinal component
of the magnetic field in the shock depends on the total magnetic
flux, $\Phi$, through the jet.  At large distances from the base
of the jet ($\theta l \gg r_j^B$, where $r_j^B$ is the jet
radius at the base of the jet), the magnetic field scales
inversely with the jet cross-sectional area
\begin{eqnarray}
B_j \approx 1.5\times 10^{-3} \Phi_{32} \theta_5^{-2} l_1^{-2},
\label{eq13}
\end{eqnarray}
\noindent
where the $\Phi_{32}$ is the magnetic flux in units of $10^{32}$ G cm$^2$.

The section of the shock farthest from the central engine (region
B in Fig.~1) seems to be the most promising location for the
production of directly observable fluxes of $\gamma$-rays in the
scenario discussed here.  The reason being that in this region
the relativistic shock is quasi-perpendicular and is likely to
accelerate particles which are strongly collimated in the shock
surface (see e.g., Kirk \& Heavens 1989, Ostrowski 1991).
Particles can be also accelerated almost rectilinearly by shock
drift acceleration (Jokipii 1987) discussed by Begelman \& Kirk
(1990).  The degree of particle collimation can be of the order
of $1/\gamma_{\parallel}$, where $\gamma_{\parallel}$ is the
Lorentz factor for the velocity of the shock measured along the
direction of the magnetic field lines. A very strong anisotropy
of accelerated particles is then possible provided that the
magnetic field is highly ordered, with no strong turbulence
present.  

The distance of the shock from the star is given by
Eq.~(\ref{eq4}) and the magnetic field at the shock can be
estimated from Eq.~(\ref{eq13}).  This enables us to estimate the
acceleration rate and the maximum electron energy allowed by
synchrotron losses.  Using Eq.~(\ref{eq15}), and Eq.~(\ref{eq13})
for the magnetic field in the jet, one finds
\begin{equation}
E_e^{\rm max} = 1.5 \times 10^6 \chi^{1/2} \Phi_{32}^{-1/2} 
\theta_5 l_1 \;\; \rm GeV.
\label{eq16}
\end{equation}
For example, for $\chi>0.04$ (expected for a perpendicular
relativistic shock) and $l_1=0.05$ one obtains $E_e^{\rm max} >
1.6 \times 10^4$ GeV.  Higher energies are possible for a highly
anisotropic particle distribution because energy losses by
synchrotron radiation depend on $(B \sin \alpha)^2$ and $\sin
\alpha$ can be small.

\subsection{Shock in the stellar wind}

In this case the shock may form as a consequence of a balance
between the jet pressure and the stellar wind pressure, or the
magnetic pressure, or as a result of a direct collision of the
jet plasma with the stellar atmosphere. The location of the shock
in first case is determined by Eq.~(\ref{eq4}). If $r_{\rm sh}
\gg R\eta^{-1}$, then the shock is quasi-perpendicular because it
is formed in the toroidal magnetic field region.  If $r_{\rm sh}
< \hspace{-1 em} ^{\sim} R\eta^{-1}$ then the shock should be
quasi-parallel since it is formed in the radial magnetic field
region.

Unless the wind shock is formed very far from the star the shock
will be non-relativistic ($v_\infty \approx 0.01 c$),
quasi-parallel, and the magnetic field in the stellar wind will
be higher than in the jet plasma. Hence, the maximum energy will
be much lower than for the jet shock, and is likely to be
determined by synchrotron losses rather than by inverse Compton
scattering.  It is therefore likely that electrons accelerated at
the wind shock may be responsible for the X-ray emission as a
result of synchrotron radiation.  To see whether this is
reasonable, we consider the energy of synchrotron photons emitted
by the highest energy electrons,
\begin{eqnarray}
\varepsilon_x \approx m_e c^2 (B/B_{\rm cr}) \gamma_{\rm max}^2,
\end{eqnarray}
where $B_{\rm cr} =4.4 \times 10^{13}$ G, and $\gamma_{\rm max}
=E_e^{\rm max}/m_e c^2$ (given by Eq.~\ref{eq15}).  For
$\varepsilon_x = 10$~keV one obtains an acceleration rate
parameter $\chi \sim 10^{-5}$ which is reasonable for a parallel
non-relativistic shock.

It is probable that the accelerated electrons may follow magnetic
field lines to region B which are likely to become directed along
the jet axis.  As the electrons in this region cool, their pitch
angles will be decrease such that their X-ray emission will be
preferentially directed in the jet direction.

\section{Pair-Compton cascade}

We consider the pair-Compton cascade initiated by relativistic
electrons accelerated at the jet shock, such that they travel
along the shock front, to estimate the spectrum of $\gamma$-rays
emerging from the AGN in the present scenario.  Since the shock
is quasi-perpendicular and relativistic we consider injection of
electrons with a power law spectrum index less than or equal to
2.  The cut-off in the electron spectrum will be determined by
the balance between particle energy gains from shock acceleration
and losses by inverse Compton scattering, and can reach
$10^4$~GeV or higher as discussed below.

\begin{figure}
\vspace{6.cm}
\includegraphics{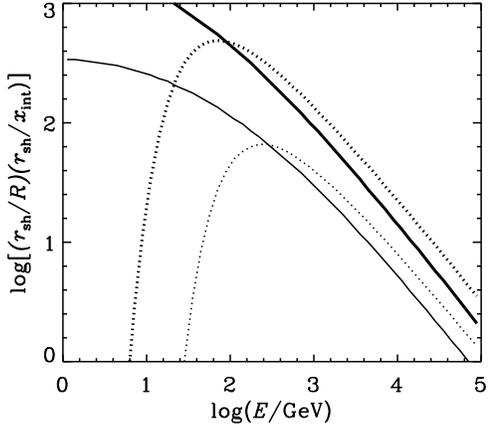}
\caption{Shock radius divided by mean free path  of electrons 
for ICS (solid curves), and shock radius divided by mean free path for 
photon-photon pair production (dotted curves), at $r=r_{\rm sh}$ multipled by 
$(r_{\rm sh}/R)$.   Lower curves correspond to $T=3 \times 10^4$~K and 
$R=10^{12}$~cm;
upper curves correspond to $T=10^5$~K and $R=2 \times 10^{11}$~cm.}
\label{fig2}
\end{figure}

We shall consider cascades in the radiation fields of OB stars
for which we adopt $T=3 \times 10^4$~K and $R=10^{12}$~cm, and
cascades in the radiation fields of Wolf-Rayet stars for which we
adopt $T= 10^5$~K and $R=2 \times10^{11}$~cm.  We have developed
a pair-Compton cascade code to calculate the emerging gamma-ray
spectrum for the case of injection of electrons at the shock
radius at the boundary between regions A and B and travelling in
the jet direction (see Fig.~1b).  This code takes account of the
anisotropic nature of the radiation field and is based on earlier
work (Protheroe 1986, Protheroe et al. 1992, Protheroe and
Biermann 1996).  Using this code, we have calculated the ratio of
the shock radius (characteristic distance scale for this problem)
to the mean free path of electrons for inverse Compton
scattering, and the ratio of the shock radius to the mean free
path for photon-photon pair production in the anisotropic
radiation field at the injection point.  This ratio, multiplied
by $(r_{\rm sh}/R)$, is plotted in Fig.~2 and we find that the
shock radius equals the mean free path for electrons of energy
$10^4$~GeV at the injection point if $r_{\rm sh} = 5.2R$ for OB
stars, or $r_{\rm sh} = 14R$ for Wolf-Rayet stars.  Using
Eq.(\ref{eq4}) we could obtain these shock radii with, for
example, the following parameters: $\dot{M}_{-5}=0.1$, $v_3=1$,
$L_{46}=0.1$, and $l_1=0.05$ (OB stars); $\dot{M}_{-5}=8$,
$v_3=1$, $L_{46}=30$, and $l_1=0.05$ (WR stars).  Clearly, these
are perfectly reasonable sets of parameters which would give rise
to pair-Compton cascading by energetic electron injection.

Finally we check whether $10^4$ GeV is a reasonable maximum
energy based on the acceleration rate and the energy loss
distance for inverse Compton scattering (given approximately by
the mean free path in the Klein-Nishina regime).  Using
Eqs.~(\ref{eq13}) and (\ref{eq14}) the acceleration distance is
given by
\begin{equation}
x_{\rm acc} = 2.2 \times 10^9 \chi^{-1} E_e \Phi_{32}^{-1} 
\theta_5^2 l_1^2 \;\; \rm cm,
\label{eq17}
\end{equation}
where $E_e$ is in GeV, giving $x_{\rm acc} = 1.4 \times 10^{12}$
cm for $\chi=0.04$, $\Phi_{32}=1$, $\theta_5=1$ and $l_1 =0.05$.
This acceleration distance is comparable to, but somewhat lower
than, the shock radii estimated above confirming that $10^4$ GeV
is a reasonable maximum energy and that the maximum energy is
probably determined by inverse Compton scattering.

We have calculated the emerging gamma-ray spectrum for a
power-law injection spectrum of the form
\begin{equation}
{dN_e \over dE_e} = \left( {E_e \over {\rm 1 \; GeV}} \right)^{-a} 
\;\; \rm GeV^{-1}
\label{eq18}
\end{equation}
for the standard shock acceleration spectrum, $a=2$, and for
$a=1.5$ representing the flatter spectra that are possible for
acceleration at relativistic shocks (Ellison et al. 1990).
Results are shown in Fig.~3(a) for OB stars and in Fig.~3(b) for
Wolf-Rayet stars for a range of shock radii spanning those
discussed above.  Specifically we show results for $r_{\rm
sh}/R=1.7$ and $r_{\rm sh}/R=17$ for OB stars, and $r_{\rm
sh}/R=4.7$ and $r_{\rm sh}/R=47$ for Wolf-Rayet stars.

\begin{figure*}
\vspace{6.cm}
\includegraphics{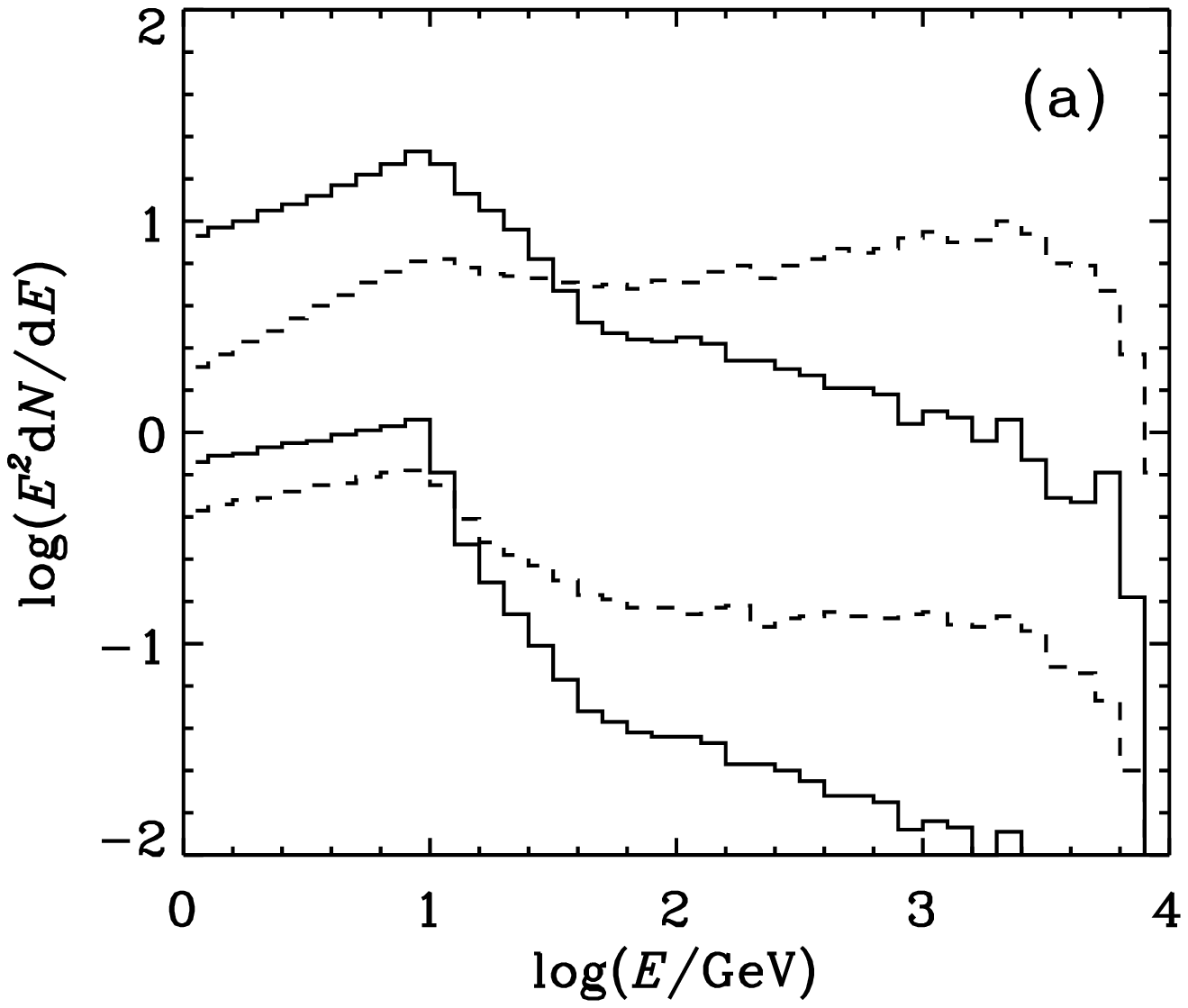}
\includegraphics{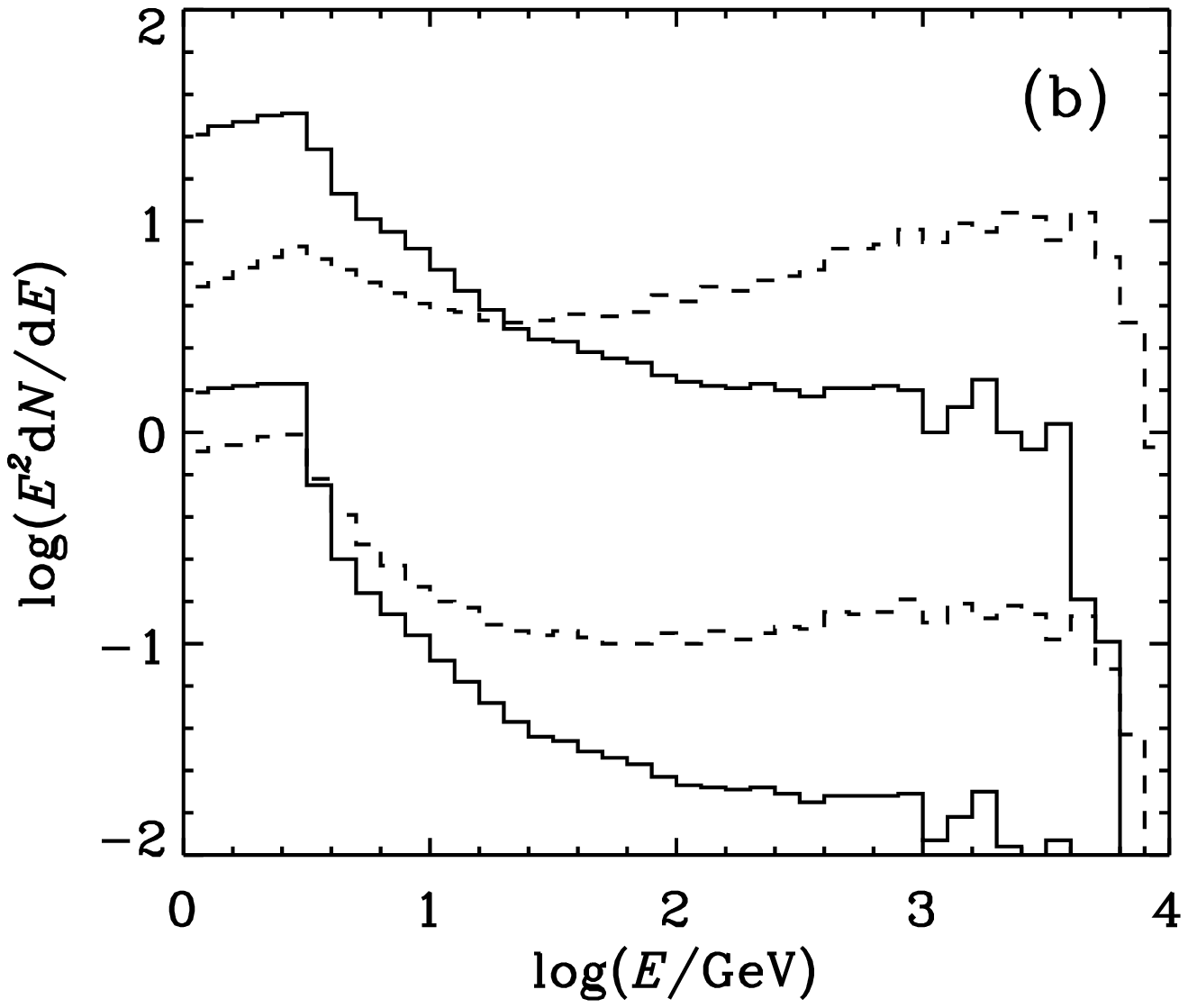}
\caption{Emerging $\gamma$-ray spectra resulting from injection of a
power-law electron spectra of the form $E^{-1.5}$ (upper curves)
and $E^{-2}$ (lower curves) extending up to $10^4$ GeV at the
shock radius in a direction tangential to the shock.  (a)
$T=3\times10^4$~K, $R=10^{12}$, $r_{\rm sh}/R=1.7$ (solid curves)
and $r_{\rm sh}/R=17$ (dashed curves); (b) $T=10^5$~K, $R=2
\times 10^{11}$, $r_{\rm sh}/R=4.7$ (solid curves) and $r_{\rm
sh}/R=47$ (dashed curves).}
\label{fig3}
\end{figure*}

From Fig.~3 we see that when the shock is located far from the
star, i.e.  $x_{\rm int}^{\rm IC}(10^4\;{\rm GeV}) > \hspace{-1.0
em} ^\sim r_{\rm sh}$, i.e., for less powerful jets and/or large $l$,
significant gamma-ray emission will occur at TeV energies.
Conversely, if the shock is near the star, i.e.  $x_{\rm
int}^{\rm IC}(10^4\;{\rm GeV}) \ll r_{\rm sh}$, the $\gamma$-ray
spectrum is significantly steepened by photon-photon pair
production in the stellar radiation field.  For a flat electron
spectrum the TeV $\gamma$-ray flux can be comparable to or
dominate the GeV $\gamma$-ray flux (see upper dashed lines in
Fig.~3) as observed during TeV $\gamma$-ray outbursts from Mrk
421 (Buckley et al. 1996).
\section{Discussion}

The expected $\gamma$-ray luminosities (if recalculated for the
isotropic case) are
\begin{eqnarray}
L_\gamma \approx 4\times 10^6\mu L_{\rm sh} \alpha_{-3}^{-2}, 
\label{eq19}
\end{eqnarray}
\noindent
where $L_{\rm sh}$ is given by Eq.~(\ref{eq5}) or (\ref{eq8}),
$\mu$ is the efficiency of conversion of jet power into gamma
rays at the shock, and $\alpha = 10^{-3}\alpha_{-3}$ rad is the
angle of the cone of $\gamma$-ray emission.  We assume that the
axis of the emission cone is directed outwards from the source of
the jet, and so the cone sweeps across the sky as the star
crosses the jet.  The expected variability time scale associated
with the line of sight to the observer passing through the cone
of emission is then
\begin{eqnarray}
t_{var}\approx 10^7 \alpha_{-3} v_{s,3}^{-1} l_1,
\label{eq20}
\end{eqnarray}
\noindent
where $v_s = 3\times 10^8 v_{s,3}$ cm s$^{-1}$ is the transverse
velocity of the star.  For $\alpha_{-3}=1, v_{s,3}=1$, and $l_1
=0.05$ this gives $t_{var}\approx 6$ days, which is comparable
with the activity period observed simultaneously from Mrk 421 in
X-rays (Takahashi et al. 1996a) and TeV $\gamma$-rays (Buckley et
al. 1996).  It is interesting that Eqs.~(\ref{eq19}) and
(\ref{eq20}) predict that for a narrower emission cone
$\gamma$-ray fluxes should be higher and variability time scales
should be shorter.

A distant observer may see emission modulated with the rotational
period of the star if the star's magnetic axis does not coincide
with its rotational axis.  In this case, the wind and magnetic
field strengths and structures should vary almost periodically in
the region of the shock with time scale given by
\begin{eqnarray}
t_{\rm rot} = 2 \pi R v_{\rm rot}^{-1} 
\approx (1- 2)\times R_{12} v_3^{-1} \;\;\;\; {\rm days}, 
\label{eq21}
\end{eqnarray}
\noindent
where $v_{\rm rot} = (0.1- 0.2)v_{\infty} = 3\times 10^8 (0.1- 0.2)v_3$
cm s$^{-1}$ is the star's rotational velocity.  This period is
comparable to the possible quasi-periodic variability detected
from Mrk 421 (Takahashi et al. 1996a).  Note also that winds from
early type stars are very unstable, and the jet plasma may
contain some irregularities. This may result in a slight change
in the location of the shock which may significantly change the
direction of a narrow cone in which $\gamma$-rays are emitted,
and may be the reason for the flickering of the $\gamma$-ray
emission on a time scale of fractions of an hour (Gaidos et
al. 1996).

Many stars are likely to be emerged in the jet at any one
time. For a stellar density of $\sim 2\times 10^5$ M$_{\odot}$
pc$^{-3}$ within $\sim 1$ pc (observed in M32, Lauer et al. 1992),
about $\sim 40$ stars with an average mass of $\sim 10$
M$_{\odot}$ would be found inside a jet with opening angle
$5^\circ$ within $\sim 1$ pc of the central engine.  The presence
of many stars inside the jet significantly increases the
probability of a distant observer being inside the $\gamma$-ray
emission cone of one of the stars.

In conclusion, we have developed a model for gamma
ray emission in AGN jets in which particle acceleration takes
place at a shock in the relativistic jet plasma produced as a
result of a massive star in the central region of the host galaxy
moving through the jet.  The $\gamma$-rays are produced in a
pair-Compton cascade initiated by accelerated electrons in the
radiation field of the star. Our model may account for the
observed GeV to TeV $\gamma$-ray spectrum and variability of
Markarian 421 and other EGRET blazars.

\section*{Acknowledgments}

W.B. thanks the Department of Physics and Mathematical Physics at
the University of Adelaide for hospitality during his visit, and
Michal Ostrowski for useful information on the shock
acceleration. R.J.P. thanks Heino Falke for a helpful discussion.
We thank Qinghuan Luo for reading the manuscript. This research
is supported by a grant from the Australian Research Council.

{}

\bsp

\label{lastpage}

\end{document}